\documentclass{bjp}

\usepackage{latexsym,revsymb}
\usepackage{amsmath,amssymb,amsfonts}
\usepackage[super,compress]{cite}
\usepackage{hyperref}
\usepackage{color}
\usepackage{graphicx}

\newcommand{\GN}{G}
\newcommand{\ac}{\bar{a}_0}
\newcommand{\vev}[1]{\langle #1 \rangle}

\begin{document}

\title{Modified Dark Matter in Galaxies and Galaxy Clusters}

\runningheads{MDM in Galaxies and Galaxy Clusters}{D.~Edmonds, D.~Farrah, D.~Minic,
Y.~J.~Ng,  T.~Takeuchi}

\begin{start}
\author{D.~Edmonds}{1}, \coauthor{D.~Farrah}{2},
\coauthor{D.~Minic}{2}, \coauthor{Y.~J.~Ng}{3}, \coauthor{T.~Takeuchi}{2}

\address{Department of Physics, The Pennsylvania State University, Hazleton}{1}

\address{Department of Physics, Center for Neutrino Physics, Virginia Tech}{2}

\address{Institute of Field Physics, Department of Physics and Astronomy, University of North Carolina, Chapel Hill}{3}

\received{29 December 2017}

\begin{Abstract}

Modified Dark Matter (MDM) is a phenomenological model of dark matter, inspired by gravitational thermodynamics, that naturally accounts for the universal acceleration constant observed in galactic rotation curve data; a critical acceleration related to the cosmological constant, $\Lambda$, appears as a phenomenological manifestation of MDM. We show that the resulting mass profiles, which are sensitve to $\Lambda$, are consistent with observations at the galactic and galaxy cluster scales. Our results suggest that dark matter mass profiles contain information about the cosmological constant in a non-trivial way.

\end{Abstract}

\PACS {95.35.+d}
\end{start}

\section[]{Introduction}

This paper is based on a talk given at the March 2017 conference of the Bahamas Advanced Study Institute and Conferences (BASIC).

There is strong evidence that the `missing mass' problem cannot be fully solved by modifying General Relativity, but will require extra sources, generally referred to as dark matter (DM)  which are non-baryonic. 
This is most evident in the formation of large-scale structures (LSS) that agree with observations\cite{davis85,whfr91,cole05,eisen05,sprin05} as well as comparisons of the observed deuterium to hydrogen ratio to that expected from Big Bang nucleosynthesis, which shows that the bulk of matter cannot be baryonic\cite{cyburt04}. 
Observations of the power spectrum of anisotropies in the cosmic microwave background (CMB) are consistent with at least most DM being non-relativistic (\textit{i.e.} cold) and diffuse\cite{smoot92,debernar00,spergel07,komat11,planck16}. 
Other evidence includes the observations of colliding galaxy clusters, which are straightforward to explain with non-baryonic cold DM\cite{clowe06} but convoluted to explain in competing models\cite{angus06,lililin13}. 

A current problem is to understand the nature of DM. The standard paradigm of a cold, diffuse, non-relativistic DM is known as the Cold Dark Matter (CDM) paradigm\cite{Peebles:1982ff, Bond:1982uy, Blumenthal:1982mv, Blumenthal:1984bp}. It is notable that CDM is not particularly restrictive with models that fit this framework; all that is required is a Weakly Interacting Massive Particle (WIMP) in which extra and independent (from the baryonic matter) degrees of freedom are described by new weakly-interacting quantum fields \cite{Jungman:1995df}, but in which (at least from the astrophysical perspective) the mass and interaction cross section of individual WIMPs are barely constrained\footnote{We will not discuss alternative dark matter scenarios that include warm dark matter models \cite{Bode:2000gq}, axions \cite{Asztalos:2009yp}, and hidden sectors \cite{Strassler:2006im} in this paper.}. 
Note that the standard cosmological model, $\Lambda$CDM, assumes that the dark energy sector is modeled with the cosmological constant $\Lambda$ (i.e. vacuum energy), and this will remain the case in our discussion of the MDM proposal.

An obvious candidate for DM is baryons that are not easily detectable by observations of photons, either in absorption or emission. 
This led to Massive Compact Halo Object (MACHO) models, in which the DM consists of brown dwarfs, neutron stars, black holes and/or other collapsed objects. However, exhaustive searches for microlensing events that would signify the presence of such objects in our Milky Way's halo turned up far too few events to make MACHOs a significant source of DM\cite{alco96,alco97,Alcock:2000ph,calchi05,tisser07}, though recent discoveries by LIGO of gravitational waves from mergers of intermediate mass black holes\cite{Abbott:2016blz,Abbott:2016nmj} have revived interest in the possibility that DM comprises primordial intermediate-mass black holes (PIMBHs).

\section{The Mass Discrepancy Acceleration Relation in Galaxies and Galaxy Clusters}

At the galactic scale, there is an intriguing relation between dark matter and baryonic matter (BM). McGaugh and collaborators have emphasized the mass-discrepancy acceleration relation (MDAR) that occurs in a sample of galaxies with very different morphologies, spanning a wide range of stellar mass.\cite{Milgrom:2007br,McGaugh:2016leg,Lelli:2017vgz} 
This correlation is expressed as a relation between the observed acceleration $a_\mathrm{obs}$ and the expected acceleration $a_\mathrm{bar}$ from baryonic matter only as
\begin{equation}
a_\mathrm{obs} \;=\; 
\dfrac{a_\mathrm{bar}}{1-e^{-\sqrt{a_\mathrm{bar}/\ac}}}\;,
\label{McGaughRelation}
\end{equation}
where the universal acceleration constant $\ac = (1.20\pm 0.02)\times 10^{-10}\mathrm{m/s^2}$, as determined by data fits. Here we have written $\ac$ rather than $a_0\approx cH_0$, which we will use in our formulation of MDM. Given that the Hubble parameter is $H_0 = (67.74\pm 0.46)\,\mathrm{km/s/Mpc}$ (see Table 4 of Ref.~\citen{Ade:2015xua})\footnote{%
That is, $H_0 = \left[(6.581\pm 0.045)\times 10^{-8}\mathrm{cm/s^2}\right]/c$.
} 
it has been noted that
\begin{equation}
\ac \;\approx\; \dfrac{cH_0}{2\pi}\;,
\label{coincidence}
\end{equation}
which suggests that the constant $a_0$ may be cosmological in origin.\cite{Milgrom:1998sy}
\\

\begin{figure*}
  \includegraphics[angle=0,width=1.0\textwidth]{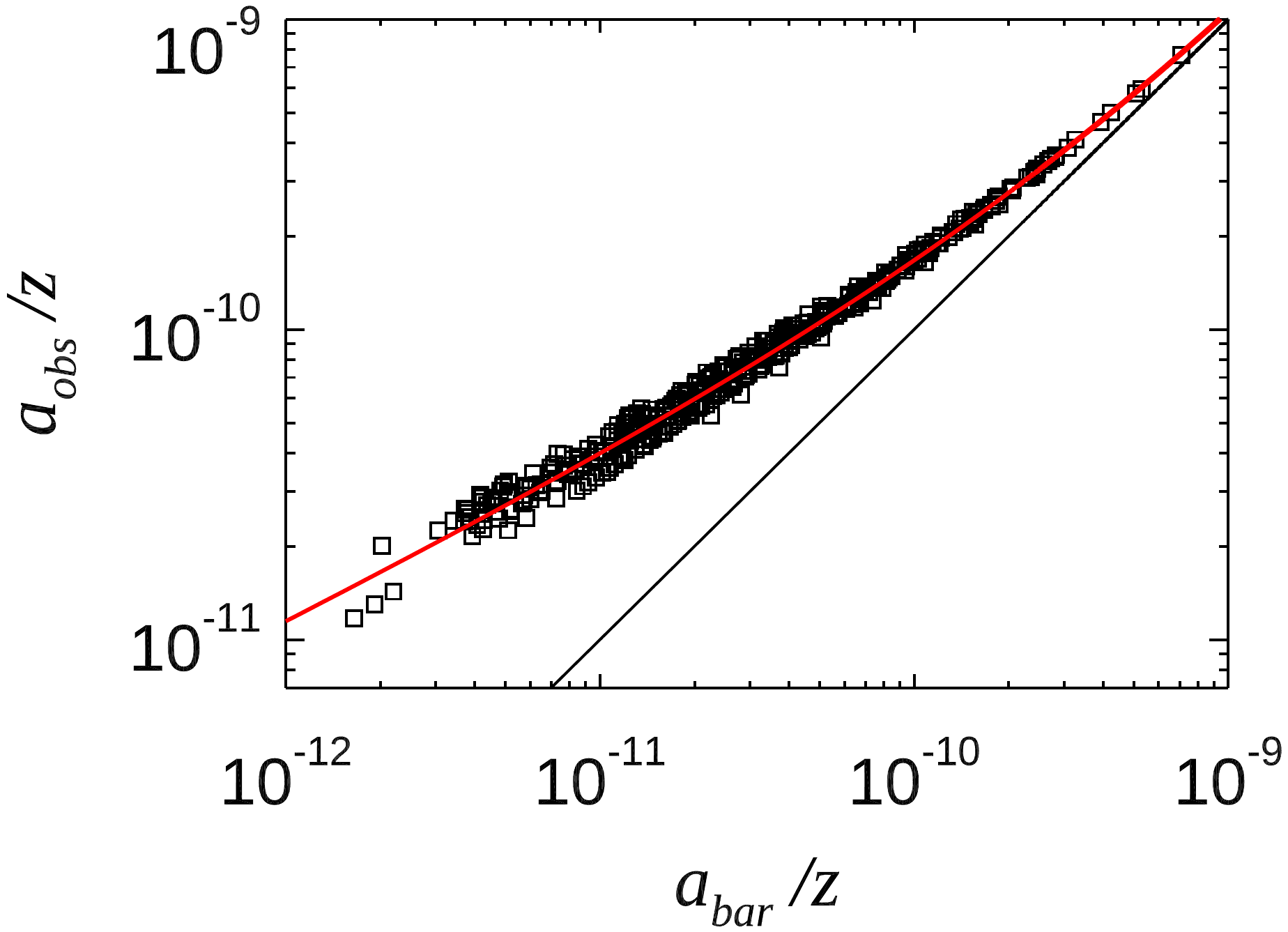}
  \caption{Comparison of observed accelerations and accelerations expected from baryons in galaxies. The black squares are 386 data points from a sample of 30 galaxies presented in this paper. The black line is what we expect from Newtonian physics and no dark matter. The red line is the prediction of Eq.~\ref{modified_McGaughRelation}.}
  \label{fig_compare_g_galaxies}
\end{figure*}

Fig.~\ref{fig_compare_g_galaxies} is similar to the one presented in Ref.~\citen{McGaugh:2016leg}, but for a different data set: We use galactic rotation curve data from the sample of Ursa Major galaxies represented in Ref.~\citen{Edmonds:2013hba}. 
The data is fit with a modified version of Eq.~(\ref{McGaughRelation}) where we introduce a scale factor 
\begin{equation}
z \;=\; \dfrac{\alpha}{\left[\,1+ \left(r/r_\mathrm{MDM}\right)\,\right]}
\end{equation}
which multiplies $\ac$ :
\begin{equation}
a_\mathrm{obs} \;=\; 
\dfrac{a_\mathrm{bar}}{1-e^{-\sqrt{a_\mathrm{bar}/z\ac}}}\;.
\label{modified_McGaughRelation}
\end{equation}
$z$ is also the prefactor which appears in Eq.~(\ref{clustermasspintro}). 
For galaxies, $z\sim 1$, and this is the same formula used in Ref.~\citen{McGaugh:2016leg}. However, inclusion of $z$ allows for consistency when we go to the galaxy cluster scale, where Eq.~(\ref{McGaughRelation}) does not fit the data well.

We would like to address the following question: Does the MDAR hold in galaxy clusters? If so, the correlation of mass profiles of DM and BM should have significance beyond coincidences of galactic dynamics within a DM halo composed of WIMPs. In other words, if the MDAR appears at different astrophysical scales, it could be pointing to yet unknown properties of the DM quanta.

\begin{figure*}
  \includegraphics[angle=0,width=1.0\textwidth]{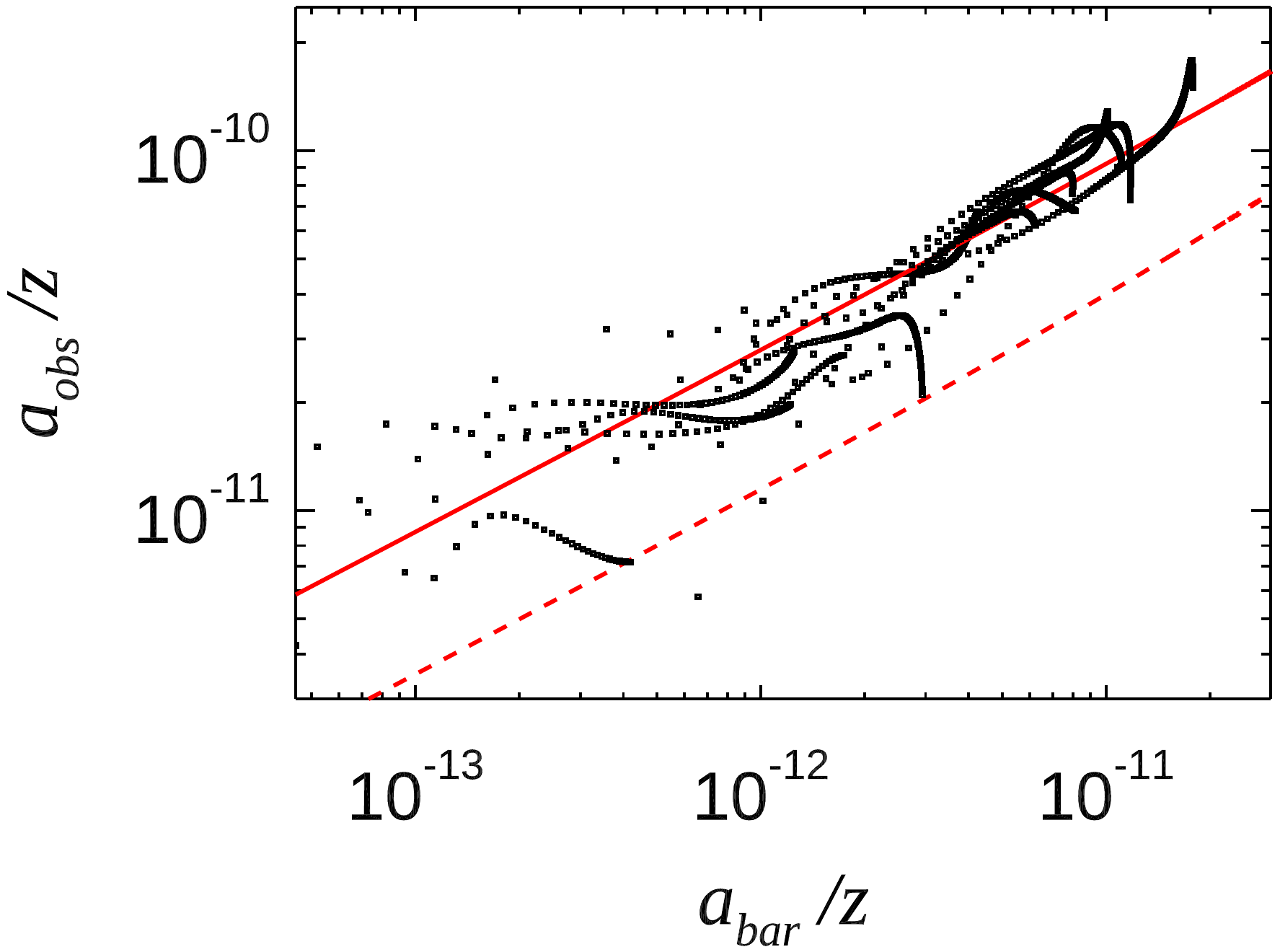}
  \caption{Comparison of observed accelerations and accelerations expected from baryons in galaxy clusters. The black squares represent fits to data from a sample of 13 galaxy clusters\cite{Vikhlinin:2005mp}. The solid and dashed red lines are the predictions of Eq.~\ref{modified_McGaughRelation} using $a_0$ and $\ac$, respectively.}
  \label{fig_compare_g_clusters}
\end{figure*}

We investigate this possibility in a sample of thirteen galaxy clusters. MDM data fits for individual clusters in this sample were presented in Ref.~\citen{Edmonds:2016tio} (see Fig.~5 for an example).
Fig.~\ref{fig_compare_g_clusters} shows a correlation beween DM and BM in the sample. The black squares represent the data fitting functions developed in Ref.~\citen{Vikhlinin:2005mp}. Our fitting function for galaxy clusters has the same form as the function used for galaxies, Eq.~(\ref{modified_McGaughRelation}), with $z$ appropriate for the galaxy cluster scale. We plot the function for two values of the acceleration scale: For the dashed red line, we use $\ac$, and for the solid red line, we replace $\ac$ in Eq.~(\ref{modified_McGaughRelation}) with $a_0$. Note that we use the same scale distance $r_\mathrm{MDM}$ for all galaxy clusters in this plot, while in the fits presented later, this scale is allowed to vary for different clusters. Using a single value increases the scatter in the data.

The MDAR for galaxy clusters is less clear than for galaxies. We do not, however, expect the correlation between DM and BM to be as tight in galaxy clusters as it is in galaxies. One of the attributes of the MDM model is that it must decouple from baryons at the cosmological scale. Exactly how this de-coupling occurs is currently under investigation. 
Since baryons should not be sensitive to $\Lambda$, 
the presence of $a_0$ in the data is expected to be due to the DM. Any de-coupling of DM and BM will therefore increase the scatter in the correlations. Furthermore, observations of galaxy clusters indicate that there are significantly fewer baryons than the cosmic baryon fraction. However, there are indications that the baryons may be pushed to larger radii.\cite{Rasheed:2010pq}. In any case, astrophysical properties of the baryonic content of galaxy clusters along with observational difficulties can easily lead to significant scatter in correlated variables.

\begin{figure*}
  \includegraphics[angle=180,width=1.0\textwidth]{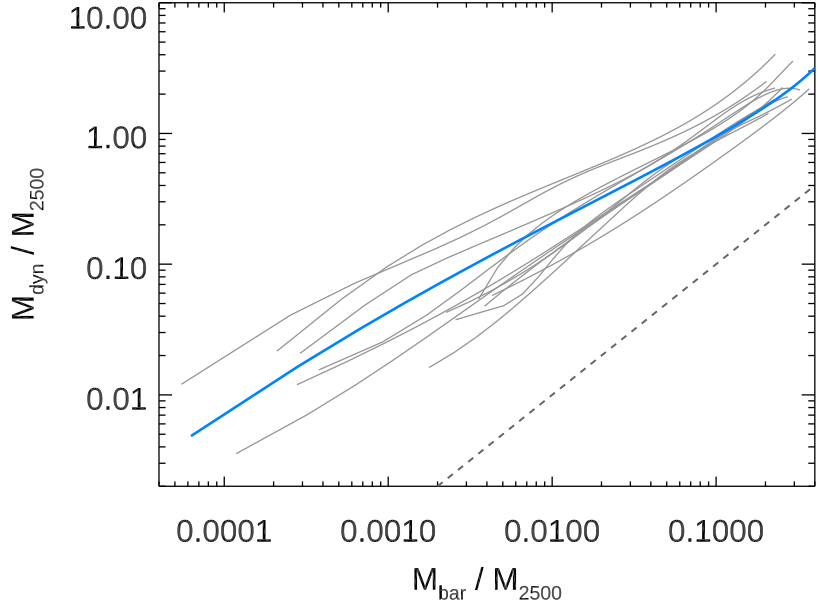}
  \caption{The solid gray lines represent the 12 galaxy clusters for which $M_{2500}$ was available. The dashed gray line is the expected relationship between dynamical and baryonic mass for Newtonian gravity and no DM. The solid blue line is the relationship expected for MDM.}
  \label{fig_compare_mass}
\end{figure*}

To better see the connection between DM and BM, we plot in Fig.~\ref{fig_compare_mass} the total mass $M_\mathrm{dyn}$ versus the baryonic mass $M_\mathrm{bar}$, both scaled to the total mass within the radius where the critical overdensity is 2500. While the correlation is not very tight, the thick, blue line, given by
\begin{equation}
\frac{M_\mathrm{dyn}}{M_{2500}} \;=\; 
\vev{z} \left\langle \frac{M_\mathrm{bar}}{M_{2500}} \right\rangle
\left( \frac{a_0}{ \vev{a} } \right)^2 ,
\end{equation}
provides a reasonable estimate for the amount of DM in a galaxy cluster for a given amount of BM, and this estimate is very sensitive to the value of $a_0$.

\begin{figure*}
  \includegraphics[angle=180,width=1.0\textwidth]{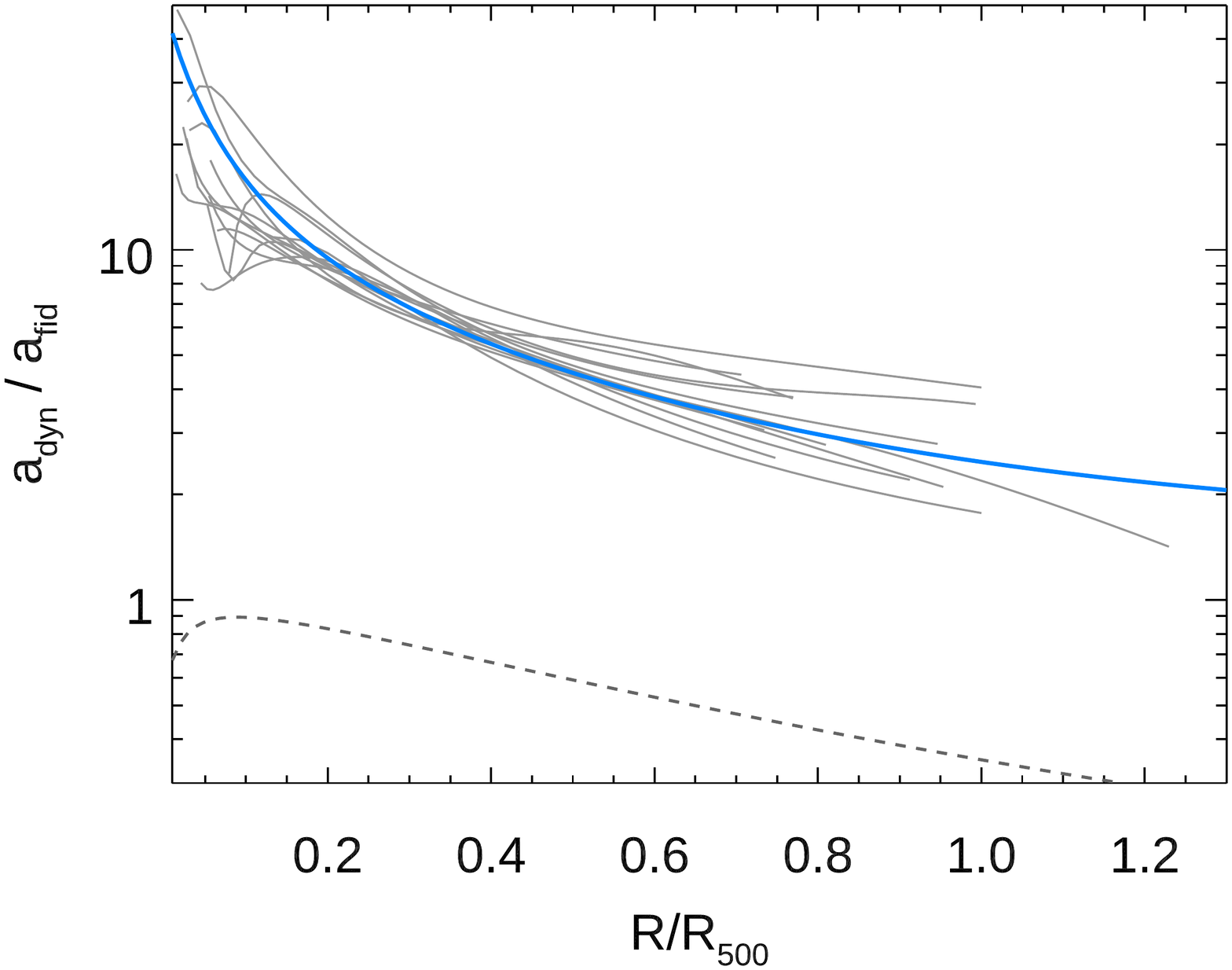}
  \caption{Colors and linestyles are the same as for Fig.~3.}
  \label{fig_acceleration_v_distance}
\end{figure*}

Another plot that illustrates the presence of $a_0$ in galaxy cluster data is presented in Fig.~\ref{fig_acceleration_v_distance}, where we plot the (scaled) acceleration as a function of distance for our sample of galaxy clusters. The thick, blue line fit to the data for this plot uses the same MDM model as used in Fig.~\ref{fig_compare_mass}:
\begin{equation}
\frac{a_\mathrm{dyn}}{a_\mathrm{fid}} 
\;=\; \vev{z}\,  
\frac{ \vev{M_\mathrm{bar} / M_{2500}} }{ \left( R/ \vev{R_{500}} \right)^2 }  
\left( \frac{a_0}{\vev{a}} \right)^2 ,
\end{equation}
where $R_{500}$ is distance from the center to the region where the critical overdensity is 500.

Notice that the above correlations are difficult to motivate within purely collision-less CDM. However, it may be possible to reconcile these results with WIMP models\cite{Ludlow:2016qzh,Navarro:2016bfs} by relying on dissipative baryonic dynamics.

On the other hand, the correlations could be pointing to yet unknown properties of DM which
connects its distribution to that of BM at galactic scales, yet allows for de-coupling at cosmological scales. Perhaps the more mysterious part of these relations is the appearance of the universal acceleration constant $a_0$. If this acceleration $a_0$  is cosmological in origin and related to $H_0$, how can the DM mass profile be sensitive to it?

\section{Further Constraining CDM Models}

In the absence of any direct detection of DM, it is important to look for other constraints that we should place on the nature and properties of DM in order to narrow down the list of possibilities. For this, we note that hints may be discernible in the tensions between observations and CDM models. We will focus on two observations at the galactic scale which strongly suggests that DM may not simply be extra, independent degrees of freedom. The aforementioned MDAR indicates a coupling between DM and BM governed by a universal acceleration scale $\ac \approx cH_0/2\pi$. The baryonic Tully-Fisher (TF) relation\cite{Tully:1977fu} is a universal relation between the total observed baryonic mass (stars + gas) of a galaxy $M_\mathrm{bar,total}$ and the asymptote of the galactic rotation curve $v_\infty$:
\begin{equation}
M_\mathrm{bar,total} \;=\; Av_\infty^4\;,\qquad
A \;=\; (47\pm 6) \,M_\odot\,\mathrm{s^4/km^4}\;.
\label{TullyFisher}
\end{equation}
This relation holds regardless of the value of $M_\mathrm{bar,total}$, or how it is distributed.
Even when the shapes of the rotation curves are different, the asymptote is always the same for galaxies with the
same baryonic mass.
This is remarkable when one recognizes that the rotation velocity $v(r)$ 
at a distance $r$ from the center of the galaxy is determined by the distribution of the
sum of baryonic and dark matter, and not by baryonic matter alone.
Nevertheless, the asymptotic velocity depends only on the total baryonic mass,
again suggesting a correlation between the BM and DM mass distributions. 

While it is not clear that current DM models  naturally explain these relations, 
they are natural consequences of Milgrom's MOdified Newtonian Dynamics (MOND).\cite{Milgrom:1983ca,Milgrom:1983pn,Milgrom:1983zz} (Note that one of the aims of our work is to provide a DM model that explains the baryonic Tully-Fisher relation and accounts for the acceleration scale $a_0$.)

In MOND, which we consider to be a law of galactic dynamics analogous to Kepler's laws for solar system dynamics, it is postulated that a simple modification of Newton's equation of motion $F=ma$ leads to flat galactic rotation curves and the TF relation without requiring DM:
\begin{equation}
F \;=\; 
\begin{cases}
ma        & (a \gg \ac) \\
ma^2/\ac  & (a \ll \ac)
\end{cases}\;.
\end{equation}
More specifically,
\begin{equation}
F \;=\; ma\,\mu(a/\ac)\;,
\label{MOND-EQM}
\end{equation}
where $\mu(x) = 1$ for $x \gg 1$ and $\mu(x) = x$ for $x \ll 1$.
The choice of interpolating functions $\mu(x)$ is arbitrary.
This implies
\begin{equation}
a_\mathrm{obs} \;=\; 
\begin{cases}
a_\mathrm{bar} & (a_\mathrm{bar} \gg \ac) \\
\sqrt{\ac a_\mathrm{bar}} & (a_\mathrm{bar} \ll \ac)
\end{cases}\;,
\end{equation}
\textit{i.e.} the same relation implied by Eq.~(\ref{McGaughRelation}).
Far away from the galactic center, we can expect the following baryonic acceleration
\begin{equation}
a_\mathrm{bar}(r) \;=\; \dfrac{\GN M_\mathrm{bar,total}}{r^2}\;,
\end{equation}
and thus
\begin{equation}
v^2(r) \;=\; r\,a_\mathrm{obs}(r) \;\xrightarrow{r\rightarrow\infty}\; r\sqrt{\ac a_\mathrm{bar}(r)}
\;=\; \sqrt{\ac\GN M_\mathrm{bar,total}} 
\;\equiv\;v_\infty^2
\;,
\end{equation}
which gives us flat rotation curves\footnote{%
In reality, rotation curves are not all flat; they display a variety of 
properties. See, e.g. Ref.~\citen{Persic:1991}.
}
and
\begin{equation}
M_\mathrm{bar,total} \;=\;
\dfrac{v_\infty^4}{\ac\GN}
\;=\; (63\,M_\odot\,\mathrm{s^4/km^4})\,v_\infty^4
\;,
\end{equation}
cf. Eq.~(\ref{TullyFisher}).
Other studies of MOND\footnote{There are also the relativistic versions
AQUAL, RAQUEL and TeVeS; but they tend to be more limited in their
predictive power.  See Ref.~\citen{Famaey:2011kh} and references therein.} 
in the context of rotation curves include
Refs.~\citen{Sanders:1996ua,Sanders:1998gr}.


We note, however, that MOND can also be interpreted as the introduction of a very specific type of DM.
Consider a spherically symmetric distribution of baryonic matter where
$M_\mathrm{bar}(r)$ is the total baryonic mass enclosed in a sphere of radius $r$.
Then, the gravitational force on a test mass $m$ placed at $r$ due to this distribution will be given by
\begin{equation}
F(r) \;=\; \dfrac{\GN M_\mathrm{bar}(r)m}{r^2}\;.
\end{equation}
Eq.~(\ref{MOND-EQM}) in this case can be rewritten as:
\begin{equation}
a(r)
\;=\;
\dfrac{1}{\mu(a(r)/\ac)}\dfrac{\GN M_\mathrm{bar}(r)}{r^2}
\;\equiv\;
\dfrac{\GN[M_\mathrm{bar}(r)+M_\mathrm{DM}(r)]}{r^2}
\;,
\end{equation}
where we identify
\begin{equation}
M_\mathrm{DM}(r) \;=\; \left[\dfrac{1}{\mu(a(r)/\ac)}-1\right]M_\mathrm{bar}(r)\;,
\end{equation}
as the total DM mass within a radius of $r$ form the center.
Thus, to reproduce the success of MOND at galactic scales, we need a DM model which 
predicts such a mass distribution. (Note that such a dark matter model is {\it not} going to be an
inversion of Milgrom's MOND, {\it i.e.} it is {\it not} going to be a  ``phantom'' dark matter, because it will
have to work on all scales: galactic, cluster and cosmological.)

\section{Modified Dark Matter}

The question regarding the relation between the fundamental acceleration parameter $a_0$ and dark matter is still outstanding. The nature of this question motivated us to examine a new model for non-baryonic dark matter, which we term modified (or ``Mondian'') dark matter, or simply, MDM. 
The idea here is that by taking into account the existence of the fundamental acceleration as well as of the baryonic TF relation, without modifying the Einstein equations, and thus Newtonian dynamics in the non-relativistic regimes, and by combining it with the non-baryonic dark matter paradigm, we should be able to sharpen the CDM proposal, and point towards a more focused origin of dark matter quanta. (At the moment, the nature of dark matter quanta is not constrained at all, and these can span enormous energy scales.)

The defining feature of the MDM proposal is that the modified dark matter profile should be sensitive to the fundamental acceleration $a_0$, or alternatively, to the cosmological constant, at all scales (galactic, cluster and cosmological) and that on galactic scales the modified dark matter mass profile should be correlated to the baryonic mass profile. Our attempt is to modify the energy momentum tensor in such a way so that the modification depends both on the original baryonic source, and on the inertial properties, such as the acceleration, associated with the geometric side of Einstein's equation.

The idea is that the acceleration can be re-interpreted in terms of temperature of the Unruh-Hawking kind \cite{Davies:1974th,Unruh:1976db}, and that in turn, such temperature can also be corrected by the presence of the cosmological constant, due to the fact that maximally symmetric spaces with positive cosmological constant, that is, asymptotically de Sitter spaces, also have a characteristic temperature associated with their cosmological horizons. This temperature can be rephrased as the fundamental acceleration. Furthermore, any excess temperature can be interpreted as excess energy, and thus as extra matter source. Thus, the fundamental origin of dark matter is tied to the thermal properties associated with gravity in the context of effective quantum field theory in curved spacetime.

The main result of our investigation can be summarized in the following formula\footnote{Details of obtaining this result are presented in Ref.~\citen{Edmonds:2016tio}} for the mass profile of non-baryonic dark matter,
 which relates the mass of the dark matter ($M'$) with the mass of the baryonic matter ($M$) via an acceleration parameter $a_0$;
\begin{equation}
\label{clustermasspintro}
\dfrac{M'}{M}
\;=\; \dfrac{\alpha}{\left[\,1+ 
\left(r/r_\mathrm{MDM}\right)\,\right]}\left[\dfrac{a_0^2}{(a_\mathrm{obs}+a_0)^2 - a_0^2}\right]
\;,
\end{equation}
where $a_\mathrm{obs}$ is the observed acceleration, $r$ the radial distance, $r_\mathrm{MDM}$ is a dark matter distance scale, and $\alpha$ is constant factor that is of order 1 for galaxies and 100 for galaxy clusters.\footnote{The value of $\alpha$ for galaxy clusters is currently not well-constrained. Values between $\sim$50 to 100 fit the data in our sample well. In this paper, we use $\alpha=50$ for our galaxy cluster fits.}
Note that for the case of galaxies $r/r_\mathrm{MDM} \to 0$, and then the mass profile reduces simply to
\begin{equation}
\dfrac{M'}{M}
\;=\; 
\dfrac{1}{2}\left[\dfrac{a_0^2}{(a_\mathrm{obs}+a_0)^2 - a_0^2}\right]
\;,
\end{equation}
where the factor of $1/2$ is the value of $\alpha$ as determined by considerations of gravitational thermodynamics (see Ref.~\citen{Edmonds:2017zhg}). Thus, this profile works both on galactic and cluster scales, and how well it works can be seen in data fits: see Fig.~\ref{fig_compare_mass} and Fig.~\ref{fig_acceleration_v_distance}. 
In previous papers, we have tested MDM predictions
with galactic rotation curves for 30 galaxies and observed mass profiles 
for 13 galactic clusters. \cite{Edmonds:2013hba,Edmonds:2016tio,Edmonds:2017zhg}. Also, based on very general arguments, MDM should work on cosmological scales where it is
consistent with the $\Lambda$CDM paradigm.\cite{Ho:2010ca}

\begin{figure}
\includegraphics[angle=0,width=0.45\textwidth]{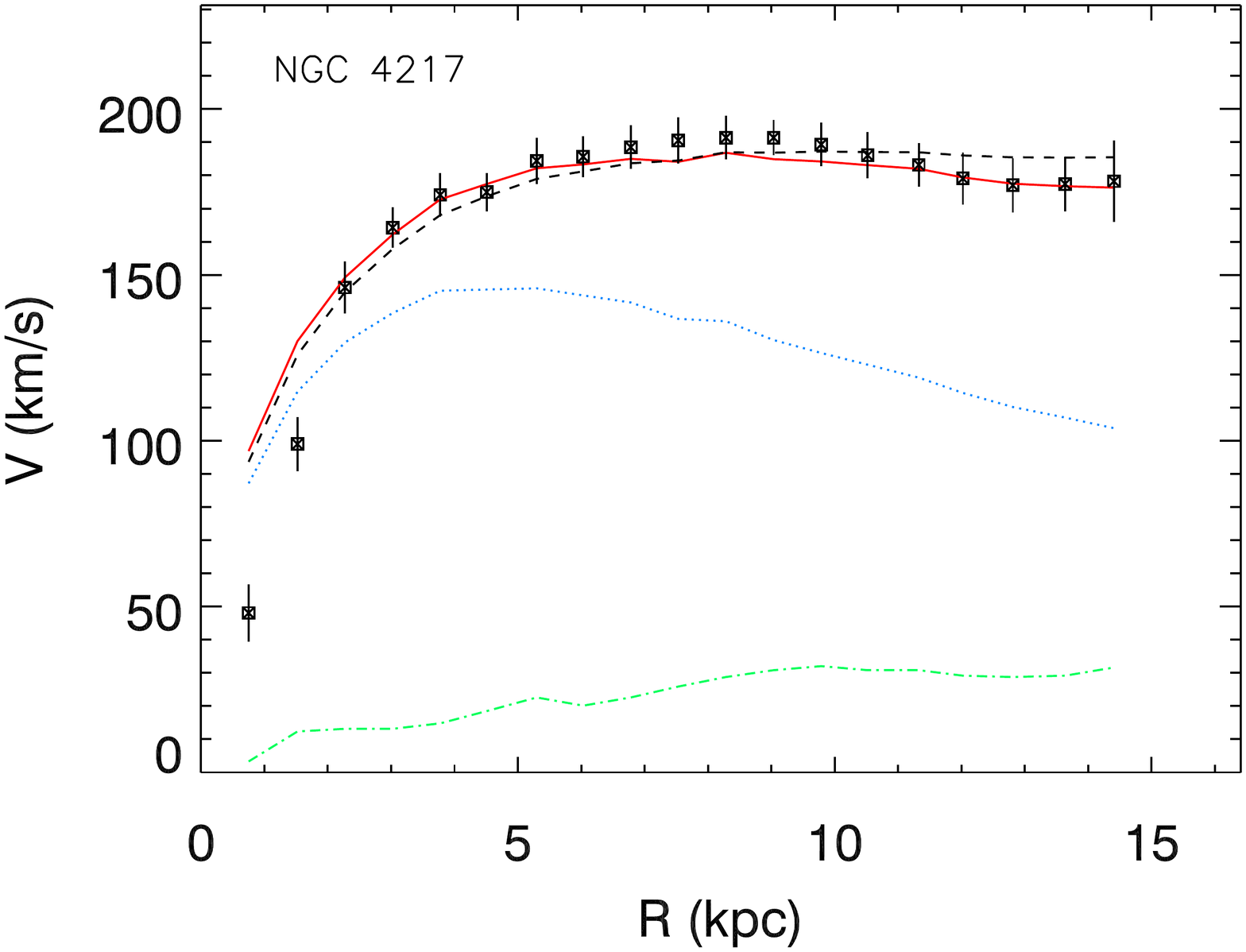} 
\includegraphics[angle=0,width=0.5\textwidth]{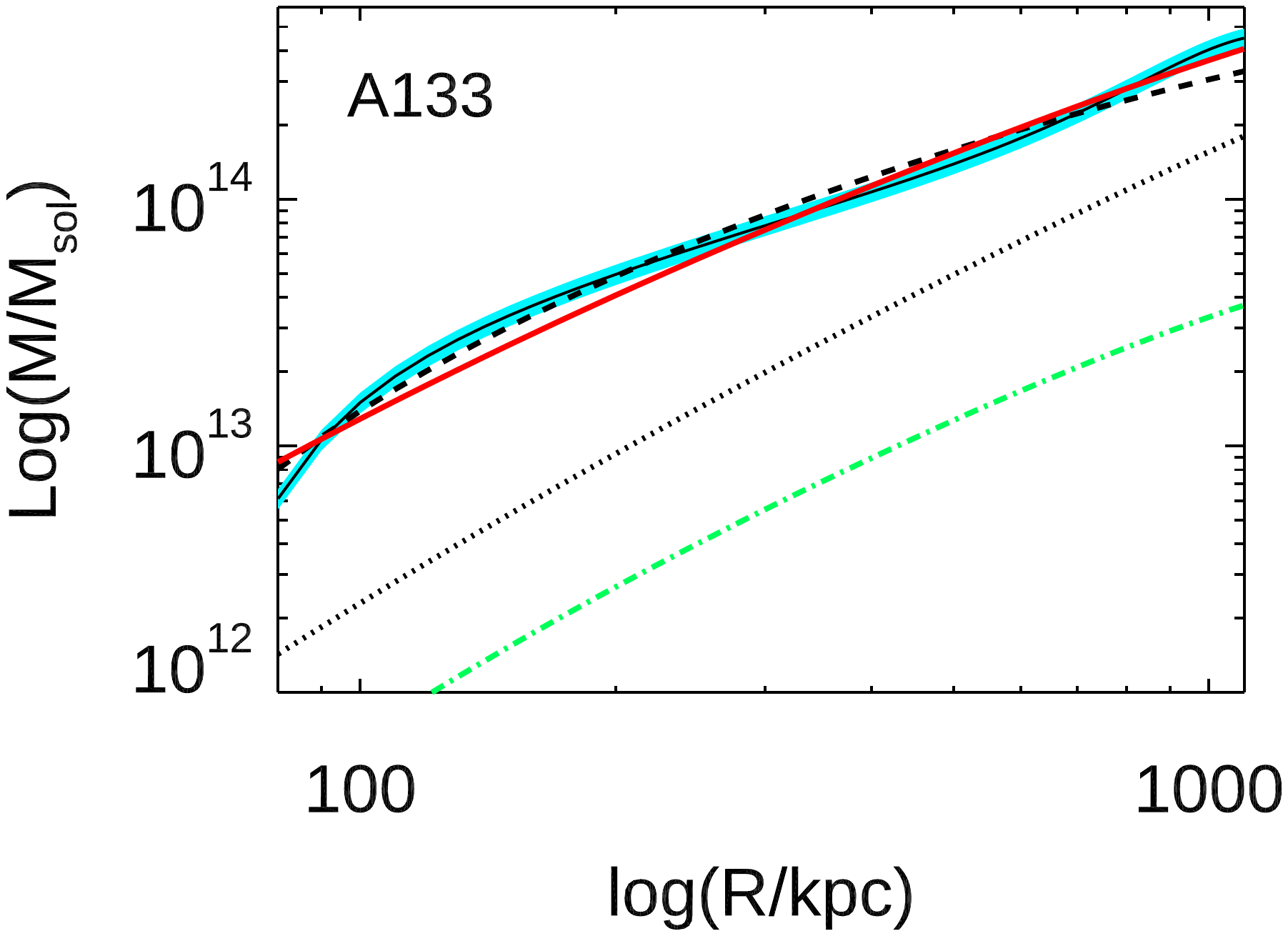} 
\\
  \caption{\textit{Left: Galactic rotation curves}. The observed rotation curve is depicted by points with error bars. The solid red and dashed black lines are the MDM and CDM rotation curves, respectively. Newtonian curves for the stellar and gas components of the baryonic matter are depicted by dotted blue and dot-dashed green lines, respectively. The mass of the stellar component is derived from the 
$M/L$ ratio determined from MDM fits to the rotation curve. \textit{Right: Galaxy Clusters} The solid black line with blue shaded region represents the measured mass profile and error bars, respectively. The green dot-dashed line is the baryonic content (gas only) and the black dotted line is the 'phantom dark matter' predicted by MOND. The black dashed line is the best-fit NFW profile, and the solid red line is the best-fit MDM model.}
  \label{fig:GRC}
\end{figure}

\section{Discussion}
Our studies of the modified dark matter mass profiles performed in the context of galaxies and the galaxy clusters reveal a characteristic acceleration that is set by the value of the Hubble constant. 
Various studies of galactic rotation curves performed by McGaugh et al.\cite{Lelli:2017vgz}, 
Frenk et al. \cite{Ludlow:2016qzh}, 
Mannheim et al. \cite{OBrien:2017bwr}, 
reveal the existence of an acceleration floor in a variety of contexts involving observational data (McGaugh), $\Lambda$CDM numerical simulations (Frenk), or modified gravity (Mannheim).\footnote{A notable modification of gravity is Moffat's MoG\cite{Moffat:2013, Moffat:2014}. To our knowledge, the acceleration floor has not been discussed in the context of MoG. It appears that MoG, in its current formulation, does not predict an acceleration floor. However, given its successes on multiple scales, it would be interesting to see if an acceleration floor emerges within the ranges of current observations.}
This acceleration floor is approximately equal to the one tenth of the characteristic acceleration set by the value of the Hubble parameter. In data, the value is set by observations of dwarf spheroidal galaxies (dSphs)\cite{Lelli:2017vgz}. In DM halo simulations that include baryonic astrophysics, the value is determined by constraints on galaxy formation.\cite{Navarro:2016bfs}
This observation is sometimes viewed as a fundamental fact (McGaugh, Mannheim) or as a coincidence found in the realm of numerical simulation (Frenk) or, possibly, in data fits (Moffat).
Here we wish to briefly point out that the value of the acceleration floor is still of the order of magnitude set by the Hubble constant, and that it is tied to the value of the cosmological constant.
Now, it turns out that the classic work of Weinberg\cite{Weinberg:1987dv} reveals a bound on the value of the cosmological constant set by the formation of gravitationally bound states, such as galaxies.
This is usually interpreted from an anthropic point of view.
However, one can view this old observation by Weinberg as an indication that there exists a bound on the acceleration parameter which is tied to the formation of gravitationally bound sates, such as galaxies, and thus, that the observed value for the acceleration floor is not coincidental, but that is set by the properties of the asymptotically de Sitter background in which we live.
This nicely meshes with our view on the emergence of the fundamental acceleration parameter from the fundamental thermodynamic properties of de Sitter space, which leads to the modified dark matter dark profile.

We plan to elaborate on this observation in  a separate publication. Here, we mention that the MDM mass profiles for galaxies predict an acceleration floor which is directly proportional to $a_0$. This is true for galaxies, but we do not expect an acceleration floor in galaxy clusters. Based on the MDM mass profiles, we expect accelerations to asymptotically fall off as $1/r$.

A striking feature of the MDM (phenomenological) model is the sensitivity of DM to different scales. Thus, MDM quanta are not particle-like, but might be modeled as metaparticles, excitations of metastrings, which look roughly like two entangled particles in momentum space.\cite{Freidel:2017nhg, Freidel:2017wst} Due to their extended (and thereby non-local) nature, such quanta could have unusual statistics, such as infinite statistics.\cite{Greenberg:1989ty, Ho:2012ar}

\section*{Acknowledgments}

We are grateful for helpful discussions with C.~Frenk, P.~Mannheim, S~McGaugh and J.~Moffat as well as participants of the BASIC~2017 conference. We thank Eduardo Guendelman and Thom Curtright for organizing this conference. We especially thank our former collaborator C.~M.~Ho for his tremendous contributions to every aspect of our work on MDM. DM thanks the Julian Schwinger Foundation for support. YJN was supported in part by the Bahnson Fund and the Kenan Professorship Research Fund of UNC.

\bibliographystyle{ws-ijmpd}
\bibliography{mdm}{}

\end{document}